%% file: main.tex
\newif\ifdraft
\draftfalse

\ifdraft
 \documentclass[draft]{jot}
\else
  \documentclass[]{jot}
\fi

\usepackage[draft]{changes}

\jotdetails{
    volume=24,      
    number=2,       
    articleno=a3,   
    year=2025,      
    license=ccbyncnd    
}

\articletype{column} 

\usepackage[T1]{fontenc}
\usepackage[utf8]{inputenc}
\usepackage{comment}

\ifdraft
 \usepackage{drafts}
\else
 \usepackage[nodrafts]{drafts}
\fi
\usepackage[norelnum,twocolumn]{figures}
\usepackage{graphicx}
  \usepackage{tcolorbox}

  \tcbset{
    colback=gray!5!white,      
    boxrule=0pt               
  }

\begin{document}

  \title{Towards a Science of Developer eXperience (DevX)}

    \author[$\ast$]{Benoit Combemale}
    \affil[$\ast$]{Inria \& University of Rennes, France}
      
\runningtitle{Towards a Science of Developer eXperience (DevX)} 

\runningauthor{Benoit Combemale}


 \input{tex/abstract}





\maketitle

\urlstyle{rm}

\section{Introduction}

In a world increasingly shaped by software, we are witnessing not only the proliferation of software products but the digital transformation of nearly every aspect of our professional and personal life. This transformation is marked by pervasive, adaptive, contextualized, and personalized digital services that have become integral to modern society. While software engineering as a discipline has matured significantly since its formal recognition at the 1968 NATO conference, the field has largely concentrated on technical challenges across the software development lifecycle. In doing so, it has often overlooked a critical dimension: the experience of developers themselves \cite{fagerholm2012developer, forsgren2024devex}. We use the term \textit{developer} broadly, to encompass not only software professionals typically identified as software engineers, but also scientists, domain experts, and even citizens (e.g., end-usesr programming \cite{BARRICELLI2019101}) who interact with software development artefacts. Their experience impacts not only the engagement in key development activities such as design \cite{palomino2025enhancing} and test \cite{parry2022surveying}, but also the overall productivity \cite{razzaq2024systematic,noda2023devex}. 

Today, as the demand for high-quality, complex software systems expands across diverse domains and involves very heterogeneous stakeholders, there is an urgent need to focus on the human aspects of software creation. 
%
While previous research has explored ways to measure \cite{d2024measuring}, understand, and improve \cite{greiler2022actionable} the experience of developers, a systematic articulation of the underlying rationales and core challenges remains lacking.
This column aims to discuss the motivations for recognizing \textit{Developer eXperience} (DevX) as a research field in its own right. 
We outline the foundational rationales for this emerging discipline, explore its intersections with existing scientific domains and key enablers, and discuss the scientific challenges. Our goal is to prompt the scientific community to acknowledge and address this vital dimension of software engineering, ultimately fostering more sustainable, effective, and inclusive software development practices.

\section{Motivations}

While addressing the complexity of software systems and meeting the required quality assurance have been the motivations for software engineering, it is important to review the current and new motivations that lead to a better consideration of the experience of engaged developers. In the following, we discuss the new dimensions of the complexity of software systems, as well as the diversity of developers and the limits of the anthropic principle in the software development lifecycle. 

\subsection{System complexity}

Modern software systems are more intricate than ever before, in the form of socio-technical eco-systems \cite{bencomo2024abstractionengineering}. They are characterized by distributed architectures, microservices, machine learning models, and large-scale integrations. Developers must reason about layers of abstractions, manage dependencies, and debug interactions across components that are often opaque or poorly documented. This complexity poses significant cognitive challenges, leading to inefficiencies, errors, and frustration. 

To succeed with the development of modern software, organizations must have the agility to adapt faster to constantly evolving environments to deliver more reliable and optimized solutions that can be adapted to the needs and environments of their stakeholders including users, customers, business, development, and IT. However, stakeholders are missing tool support for global decision-making, considering the increasing variability of the solution space, the frequent lack of explicit representation of its associated variability and decision points, and the uncertainty of the impact of decisions on stakeholders and the solution space  \cite{kienzle:hal-03770004}. This leads to an ad-hoc decision making process that is slow, error-prone, and often favors local knowledge over global, organization-wide objectives. It urges to provide automation and tool support in aid of a multi-criteria decision making process involving different stakeholders within a feedback-driven software development process where feedback cycles aim to reduce uncertainty.

While software is revolutionizing the modern world, software systems evolve under frequently changing environments, and are expected to handle  ever-increasing uncertainty. This blurs not only the line between engineering-time and execution-time \cite{10.1145/1882362.1882367}, but also  between software and the real world as both are fusing into a single fabric. These dynamics require accelerated levels of adaptability---indeed, a \emph{temporal} adaptability, i.e., the ability to adapt not only to a fixed space of variable requirements, but also to an emerging chain of changing requirements, often driven by incoming input data.

\subsection{Developer diversity}

The developer community is no longer limited to traditional software engineers. Scientists, domain experts, and even non-programmers—often referred to as "citizen developers"—increasingly engage in programming to advance their fields or solve domain-specific problems. This heterogeneity creates challenges in designing tools and processes that cater to diverse skill levels, mental models, and goals. For example, while experienced software engineers may seek extensibility and fine-grained control, citizen developers prioritize simplicity and intuitive interfaces \cite{bucaioni2022modelling}, domain experts expect to manipulate relevant abstractions of the domain to bridge the gap between the problem space and the solution space \cite{madni2018model}, and scientists relies on complex processes and various software languages to move from continuous mathematical models to efficient implementation on HPC infrastructures \cite{9622288}. 

\subsection{Complexity of Software Development Practices}

Modern software development has evolved into a highly intricate discipline characterized by a multiplication of processes, tools, and responsibilities. Far from being limited to coding alone, contemporary development practices encompass various tasks, including requirement engineering, architectural design, implementation, testing, deployment, infrastructure management, monitoring, and continuous maintenance and evolution. Each of these dimensions introduces its own set of artifacts and tools, leading to increasing cognitive and operational overhead for developers.

A major contributor to this complexity is the proliferation of development artifacts—ranging from source code and documentation to configuration files, test scripts, CI/CD pipelines, and telemetry data. These artifacts are often managed across heterogeneous tools such as version control systems, issue trackers, integrated development environments (IDEs), build servers, and cloud platforms. The resulting fragmentation demands frequent context switching, which can negatively impact developer productivity and mental well-being \cite{10.1145/3454122.3454124,meyer2014software}.

Moreover, developers are now expected to assume responsibilities traditionally handled by specialized roles. Practices like DevOps and Site Reliability Engineering (SRE) blur the lines between development, and operations. This convergence, while promoting agility and collaboration, also necessitates continuous learning and adaptation to new frameworks, paradigms, and toolchains.

The pace of technological change further exacerbates the complexity. The rapid evolution of programming languages, libraries, cloud-native technologies, and architectural styles (e.g., microservices, serverless computing) demands that developers not only build software but also stay perpetually current with trends and best practices.

Hence, the complexity of software development today arises not just from the systems being built but also from the ecosystem in which they are developed, thus degrading the developer experience.

\section{Rationales for DevX}

In an era where software systems underpin critical infrastructures and drive global innovation, the experience of developers —the creators of these systems— has become a central concern. Developer eXperience (DevX) transcends traditional software engineering metrics by emphasizing the holistic impact of tools, workflows, and environments on developers' creativity, engagement, and overall effectiveness \cite{forsgren2024devex}. DevX is not merely a productivity enhancer; it is an enabler of accessible, pleasurable, and accountable development practices that align with the growing complexity and diversity of modern software ecosystems. In the following, we review the rationales of DevX from the point of view of software engineering. 

\subsection{Fostering Creativity and Engagement}

Creativity is a cornerstone of effective software development. Developers constantly devise novel solutions to technical challenges, a process that thrives in environments fostering engagement and flow. Poorly designed tools, fragmented workflows, and opaque systems disrupt this creative process, eroding engagement and increasing cognitive load. DevX research highlights the importance of affordances—features that intuitively guide developers toward productive interactions. By enabling seamless exploration, experimentation, and debugging, a strong DevX cultivates environments where developers can fully leverage their creative potential.

\subsection{Enhancing Accessibility and Confidence}

The democratization of software development has brought diverse actors into the field, including scientists, domain experts, and citizen developers. Accessibility is a critical aspect of DevX, ensuring that tools and processes cater to varying levels of expertise and technical backgrounds. Equally vital is fostering confidence, as intuitive and transparent systems help developers trust their tools and outputs. By prioritizing accessibility and confidence, DevX bridges the gap between novice and expert developers, empowering all to contribute effectively.

\subsection{Prioritizing Social Translucence and Accountability}

Modern software development is inherently collaborative, often involving distributed teams working across different time zones and roles. Social translucence—the ability to make others' actions and intentions visible in a respectful and comprehensible way—is a key attribute of effective development environments. Tools like forges and practices like DevOps promote accountability by offering clear records of contributions while facilitating open communication and feedback. DevX research emphasizes the role of such systems in building trust, reducing misunderstandings, and ensuring that collaborative processes are as seamless as individual workflows.

\subsection{Balancing Pleasure with Productivity and Efficiency}

A compelling DevX is not merely functional; it is also pleasurable. Developers derive satisfaction from smooth workflows, elegant tools, and systems that anticipate their needs. This pleasure, in turn, feeds motivation and engagement, fostering sustainable productivity. For instance, tools that offer visual feedback, gamified elements, or aesthetic design can transform routine tasks into enjoyable experiences. However, pleasure must align with productivity and efficiency. Tools should minimize repetitive or redundant work and allow developers to achieve their goals swiftly without sacrificing the joy of problem-solving. Striking this balance is a central challenge in DevX research.

\subsection{Addressing Overwhelming Complexity and Artifact Management}

As software systems grow in complexity, developers must navigate a large set of software development artefacts—source code, documentation, tests, logs, configurations—across numerous tools and platforms. This fragmentation often leads to inefficiencies, as developers are forced to context-switch or manually integrate disparate workflows. DevX research seeks to streamline these processes, offering integrated environments that provide developers with a clear mental model of their work and reduce friction. By improving efficiency and productivity, such environments enable developers to spend more time on creative and high-impact tasks.

\section{Key Enablers and Scientific Challenges}

The experience of software developers is a multifaceted concern that extends beyond the technical concerns related to artefacts, tools, and methodologies. It also encompasses human and social dimensions, thereby demanding a comprehensive perspective. In this section, we identify key enablers and discuss the remaining open scientific challenges, with attention to both technical and social aspects.

\subsection{Technical Dimension}

From a technical standpoint, enhancing the developer experience hinges on the evolution of programming environments, languages, and tools. One fundamental concern is program comprehension. As systems grow in complexity, understanding their structure and behavior becomes increasingly difficult. This necessitates improved static and dynamic analysis techniques, better visualization methods, and deeper insights into the cognitive processes involved in reading and modifying code.

The interfaces through which developers interact with code also play a critical role. Programming notebooks, such as Jupyter, support exploratory and data-centric workflows, yet still present usability and reproducibility challenges. Similarly, diverse programming styles—ranging from live and exploratory programming to literate programming—demand flexible and responsive environments that support rapid feedback, inline documentation, and iterative development. Methods to facilitate the understanding of the overall behavior—such as example-based programming \cite{10.1145/3426428.3426919}, omniscient debugging \cite{5287015} and moldable development \cite{10.1145/3698322.3698327}—offer promising avenues but require further refinement and integration with the various programming styles and interfaces.

Another important area involves the design and implementation of program representations that are both abstract and manipulable. These representations should enable intuitive interaction, particularly for end-users. Visual programming environments (incl., block-based programming environment \cite{weintrop2017comparing}) attempt to address this by accelerating the learning curve, though issues related to scalability, expressiveness, and debugging remain open.

To better support a wide range of users and tasks, programming languages and environments are increasingly incorporating mixed notations that combine text, graphics, and symbolic elements. Creating coherent semantics and robust tooling around these hybrid notations presents a significant research challenge. Closely related is the field of abstraction engineering, which focuses on developing, managing, and evolving meaningful abstractions. 

As software development processes shift toward continuous practices, such as continuous integration and deployment, new demands are placed on tools and infrastructures. Environments must support real-time feedback and ongoing changes while maintaining stability. This, in turn, drives the need for more adaptable input and output modalities—including touch, voice, and gesture-based interactions—that can make programming more accessible and intuitive.

Finally, there remains a strong need for infrastructures, meta-tools, and frameworks that streamline the creation and extension of programming environments. These tools are essential not only for researchers and educators but also for practitioners who wish to customize or prototype new workflows.

\subsection{Social Dimension}

Beyond the technical realm, the social dimension of developer experience introduces a range of human-centric and interdisciplinary challenges. Addressing these effectively requires the engagement of researchers from the social sciences and humanities (SHS), alongside computer scientists and software engineers.

At the individual level, it is important to consider how concepts such as affordance, cognitive load, motivation, and emotional engagement influence the way users interact with programming environments. Tools should be designed with these psychological and ergonomic factors in mind, to foster inclusive and empowering experiences for diverse populations.

At the collective level, social dynamics such as collaboration, communication, and shared understanding come to the forefront. Concepts like social translucence—the visibility of others’ activities and intentions—are vital for supporting teamwork and community practices. Designing programming environments that promote awareness, coordination, and mutual support among users remains an open challenge.

In summary, the development of future programming environments must be informed by both technical innovation and human-centered design. Progress in this area will depend on a sustained commitment to cross-disciplinary research and on tools that reflect the complexity of both the artefacts and the communities that produce and maintain them.        

\section{Conclusion}

Developer experience (DevX) is a multidimensional construct that intertwines creativity, engagement, accessibility, confidence, and collaboration. By focusing on affordance, social translucence, accountability, and the interplay of pleasure with productivity and efficiency, DevX research addresses the core challenges of modern software engineering. As systems grow more complex and developers more diverse, a systematic investment in DevX is essential to empower any developer while ensuring a sustainable evolution of software development practices.

\input{tex/acknowledgement}

\input{tex/about}

\bibliography{bib/biblio}

\end{document}

%% file: tex/abstract.tex
 \begin{abstract}
As software continues to permeate nearly every facet of modern life, the complexity and ubiquity of digital services underscore the need for sustainable, effective, and inclusive software development practices. Although software engineering has made significant progress in technical challenges since its inception, the human experience of those involved in software creation, broadly defined as developers, remains underexplored. This column advocates for the formal recognition of Developer eXperience (DevX) as a distinct research field. We argue that DevX profoundly influences critical development activities and overall productivity, especially as development becomes increasingly collaborative and diverse in terms of application domains. Building on existing efforts to measure and enhance DevX, we identify key rationales, scientific enablers, and interdisciplinary intersections that support this emerging discipline. We also outline the core scientific challenges ahead, aiming to call for actions from the research community and to promote more human-centered approaches to software engineering.
 \end{abstract}

%% file: tex/acknowledgement.tex
\acknowledgment{
The author thanks all early readers of this text for their valuable feedback. I also kindly thank Valentin Bourcier and Steven Costiou for our insightful discussions, as well as all the partners that shared their points of view during a workshop in France on December 2024 dedicated to the topic of developer experience. 
}

%% file: tex/about.tex
\section*{About the authors}
\shortbio{Benoit Combemale}{ is a Research Director at Inria and a Full Professor of Software Engineering at the University of Rennes. His research interests in Software Engineering include Software Language Engineering, Model-Driven Engineering, and Software Validation \& Verification. 
\authorcontact[http://combemale.fr]{benoit.combemale@inria.fr}}